\documentclass[journal]{IEEEtran}

\usepackage[dvips]{color}
\usepackage{graphicx}
\usepackage{amsmath}

\begin{document}
\title{Cognitive Network Cooperation for Green Cellular Networks}
\markboth{IEEE Access (accepted to appear)}%
{Jia Zhu \MakeLowercase{\textit{et al.}}: Cognitive Network Cooperation for Green Cellular Networks}

\author{Jia Zhu and Yulong~Zou,~\IEEEmembership{Senior Member,~IEEE}

\thanks{This work was partially supported by the National Natural Science Foundation of China (Grant Nos. 61302104, 61401223 and 61522109) and the Natural Science Foundation of Jiangsu Province (Grant Nos. BK20140887 and BK20150040).}
\thanks{J. Zhu is with the School of Telecommunications and Information Engineering, Nanjing University of Posts and Telecomm., China. (Email: jiazhu@njupt.edu.cn)}
\thanks{Y. Zou (corresponding author) is with the School of Telecommunications and Information Engineering, Nanjing University of Posts and Telecommunications, Nanjing 210003, Jiangsu, P. R. China. (Email: yulong.zou@njupt.edu.cn)}

}

\maketitle

\begin{abstract}
In recent years, there has been a growing interest in green cellular networks for the sake of reducing the energy dissipated by communications and networking devices, including the base stations (BSs) and battery-powered user terminals (UTs). This paper investigates the joint employment of cognition and cooperation techniques invoked for improving the energy efficiency of cellular networks. To be specific, the cellular devices first have to identify the unused spectral bands (known as spectrum holes) using their spectrum sensing functionality. Then, they cooperate for exploiting the detected spectrum holes to support energy-efficient cellular communications. Considering the fact that contemporary terminals (e.g., smart phones) support various wireless access interfaces, we exploit either the Bluetooth or the Wi-Fi network operating within the spectrum holes for supporting cellular communications with the intention of achieving energy savings. This approach is termed as \emph{cognitive network cooperation}, since different wireless access networks cognitively cooperate with cellular networks. In order to illustrate the energy efficiency benefits of using both cognition and cooperation, we study the cooperation between television stations (TVSs) and BSs in transmitting to UTs relying on an opportunistic exploitation of the TV spectrum, where the unused TV spectral band is utilized in an opportunistic way, depending on whether it is detected to be idle (or not). It is shown that for a given number of information bits to be transmitted, the total energy consumed is significantly reduced, when both cognition and cooperation are supported in cellular networks, as compared to the conventional direct transmission, pure cognition and pure cooperation.
\end{abstract}


\IEEEpeerreviewmaketitle

\section{Introduction}

\IEEEPARstart {T}{he} past few decades have witnessed an explosive growth in the development and implementation of ever more sophisticated wireless networks. The rapid growth in wireless communications has resulted in a significant increase of the energy consumption of both the service providers and mobile users. The image and video applications (e.g., video on demand, video games, etc.) rapidly drain the battery charge of mobile terminals. Specifically, the second-generation (2G) mobile devices have a relatively low power consumption of around 1-2 watts, while the third-generation (3G) terminals double this figure. The fourth-generation (4G) terminals are expected to have another twofold increase over the 3G devices' power consumption. As discussed in [1], if no action is taken to reduce the power consumption of mobile terminals, 4G terminals would be limited to operate only in certain places, or would even become restricted to locations, where power charging outlets are available, which is referred to as an ``energy trap". It is therefore extremely important to explore energy-efficient wireless communication technologies for reducing the energy consumption of both the base stations and of user terminals. At the time of writing, cognition and cooperation are envisioned to be the key enabling techniques conceived for green wireless communications.

The cognitive radio (CR) concept is emerging as one of the most promising techniques in wireless communications for addressing the spectrum scarcity by enabling dynamic spectrum access. As discussed in [2], the CR transmission process is generally composed of two essential phases: 1) the spectrum sensing phase, where CR users attempt to identify the unused spectrum (referred to as a spectrum hole) in, e.g., TV frequency bands, and 2) the cognitive transmission phase, where CR users transmit their data over the spectrum holes detected. Hence, extensive research efforts have been devoted to the CR concept, most of which are focused on spectrum sensing, access, allocation and sharing [3], [4], aiming for improving the spectral efficiency, rather than the energy efficiency. In [5], Shannon proved that the channel capacity is linearly proportional to the bandwidth, but only logarithmically to the transmit power. Having said this, it can also be readily shown that as the bandwidth tends to infinity, the capacity becomes also an increasing function of the power. This implies that given a certain target capacity, the use of additional spectral bandwidth can help reduce the transmit power in order to maintain a given channel capacity, demonstrating the classic tradeoff between the power- and bandwidth- efficiency. It has been reported in [6] that almost half of the total energy can be saved, if mobile network operators employ CR techniques for dynamically managing their licensed spectrum bands and fully exploiting the available radio resources for energy efficiency improvements. As a consequence, CR is inevitably becoming one of the key techniques of future green wireless networks.

Cooperation (also termed as cooperative communications) is also well recognized as a hot research topic in wireless communications [7]-[9]. Cooperative communications allows the distributed network nodes to assist each other for maximizing their respective interests, which provides a new perspective on wireless network optimization. This line of work was pioneered by Laneman in [7], where several cooperative relaying protocols (i.e., fixed relaying, selective relaying, and incremental relaying) were studied in terms of both the achievable rate and outage probability. In [8], the employment of space-time coding in cooperative wireless networks was further examined, showing a significant improvement in outage performance through user cooperation. It was demonstrated in [9] that cooperative techniques are capable of improving the transmission reliability with the aid of their spatial diversity gain and/or increasing the system throughput through resource aggregation. At the time of writing, cooperative communication architectures have already been adopted in several wireless networking standards, e.g., IEEE 802.16j and the long term evolution (LTE)-advanced systems. In addition to improving the reliability and throughput, cooperative communications also offer potential energy savings in wireless networks. For example, in cellular networks, if a user lies at the edge of its associated cell (a so-called cell-edge user), typically a high transmit power is required to maintain the target quality-of-service (QoS) requirement in the uplink (UL), which would significantly drain the user's battery energy and may additionally cause severe co-channel interference as well. In this case, selecting an appropriate relay to assist the cell-edge user's data transmission is capable of effectively reducing its power consumption. For example, having a perfect-reconstruction decode-and-forward relay halfway between the BS and MS in the case of an inverse fourth-power law reduces the required power by 12dB, i.e., over an order of magnitude. Therefore, it is of high practical interest to explore cooperative techniques for energy savings in cellular networks, especially for the cell-edge users.

In general, today's wireless devices (e.g., smart phones) are equipped with multiple network access interfaces, such as Bluetooth, Wi-Fi, and LTE, where the different wireless access networks involve different radio characteristics in terms of their coverage area and energy consumption. To be specific, both Bluetooth and Wi-Fi provide local area coverage at a low energy consumption, whereas LTE offers a wider coverage, but unfortunately at a higher energy consumption. Motivated by the observation that different wireless networks complement each other, we have investigated the cooperation between multiple heterogeneous wireless networks (e.g., Bluetooth and LTE) in [10], which may be referred to as \emph{network cooperation}. More specifically, we have examined the use of Bluetooth to assist LTE transmissions for improving the energy efficiency of cellular communications. It was shown in [10] that given the specific target data rate and outage probability requirements, network cooperation has the potential of significantly reducing the total energy consumption of cellular transmissions, especially when the mobile stations (MSs) roam at the edge of a cell. However, no cognitive features were considered for network cooperation in [10], leaving unused spectral bands that could otherwise be fully exploited for beneficial energy efficiency improvements. In this paper, motivated by these considerations, we address the integration of CR features into the above-mentioned network cooperation framework, which is termed to as \emph{\textbf{cognitive network cooperation}}. In the proposed framework, cellular devices are first allowed to identify spectrum holes through spectrum sensing. Then, network cooperation is invoked for efficiently exploiting the available spectrum holes for green communications.

The benefits of cognition have been investigated for cellular communications in [11], where the overlay approach was invoked for spectrum sharing between a TV broadcast system and a cellular system. The overlay-based spectrum sharing was shown to be beneficial in terms of increasing the capacity of both the downlink and uplink of cellular communications. In [12], the cognitive technique was also studied in cellular networks for intelligently allocating network resources to mitigate the effects of co-channel interference. Furthermore, in [13], the cooperation among multiple radio access technologies (RAT) was examined in the context of heterogeneous wireless networks. It was shown in [13] that the multitier and multi-RAT deployment is capable of boosting the network coverage as compared to single-tier cellular systems. Although the cognition and cooperation techniques have been studied in cellular networks [11]-[13], these research efforts were focused on the enhancement of the network capacity and coverage, instead of improving the wireless energy-efficiency. More recently, in [14] and [15], energy harvesting (EH) was invoked for reducing the power consumption of wireless sensor networks. In contrast to the network capacity improvement methods of [11]-[13], we intend to exploit the joint benefits of cognition and cooperation for the sake of enhancing the energy-efficiency of cellular communications.

The main contributions of this paper are summarized as follows. Firstly, we propose a cognitive network cooperation framework, where different wireless networks cognitively cooperate with each other for the sake of improving the achievable energy-efficiency of wireless communications. Secondly, we present the joint cognition and cooperation aided cellular communications philosophy and quantify its energy saving benefits compared to both the traditional direct transmission as well as to the pure cognition and pure cooperation. Finally, we reveal the tradeoff between the energy efficiency and outage performance, demonstrating that the energy efficiency of cellular communications improves, as the outage performance degrades.

The rest of this paper is organized as follows. Section II briefly reviews the spectrum sensing approaches conceived for detecting spectrum holes in an energy efficient manner. In Section III, we discuss the cognitive network cooperation framework invoked for efficiently utilizing the spectrum holes for energy reduction in cellular networks. Next, Section IV presents the case study of an energy efficient cellular network design and shows that the energy efficiency of cellular communications can be significantly improved by exploiting the concurrent use of cognition and cooperation. Finally, we provide some concluding remarks in Section V.

\section{Spectrum Sensing Relying on Cognition}
\begin{table*}
  \centering
  \caption{Comparison among the energy detector, matched filter detector, and feature extraction detector.}
  {\includegraphics[scale=0.7]{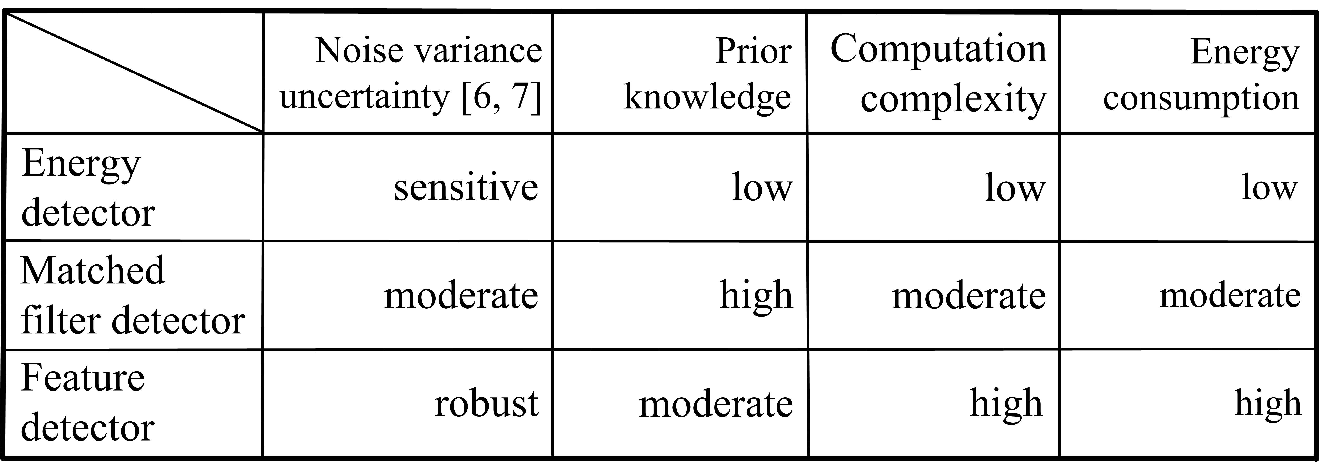}\label{Tab1}}
\end{table*}

This section is mainly focused on the identification of spectrum holes via cognitive functionalities. According to the Shannon capacity of a bandlimited noisy channel [5], there is a tradeoff between bandwidth- and power- efficiency, implying that increasing the transmission bandwidth is capable of reducing the transmit power without degrading the channel capacity. Meanwhile, the relevant studies have shown that the occupation of the licensed spectrum typically varies in the range of 15\% to 85\%, thus a large portion of the spectrum is under-utilized [2]. In particular, some frequency bands, such as the TV bands, are largely unoccupied by licensed users at certain geographic locations and particular time intervals. Hence, it is beneficial to identify and exploit the unused spectrum in these bands for cellular communications {{to reduce}} the energy consumption, while satisfying both the throughput and QoS requirements. There are two typical spectrum opportunities, namely the \emph{temporal} and \emph{spatial} spectrum opportunities. To be specific, if a spectral band is not occupied by licensed (or primary) users at a particular time, we can temporarily reuse it for unlicensed (or secondary) users, which is referred to as a \emph{temporal spectral opportunity}. By contrast, if the licensed and unlicensed users are sufficiently far away from each other, so that no excessive interference will be imposed when they transmit over the same frequency band, the unlicensed users can fully reuse the licensed users' spectral band, which may be termed as a \emph{spatial spectral opportunity} [16], [17].

In the recent literature on spectrum sensing, three different types of signal detection methods have been considered for the identification of spectrum holes, namely energy detection (ED), matched filtering (MF), and feature extraction (FE) [3]. In general, an energy detector accumulates the energy of the signal received over a given frequency band and then compares it to a predefined threshold to decide whether the spectral band is occupied by licensed users or not. To be specific, if the accumulated energy is lower than the predefined threshold, the observed spectrum band is deemed to be idle; otherwise, the band is regarded as being occupied by the licensed user. We note that ED is unable to readily differentiate the desired signal from both the interference and the noise, hence it is prone to missed detection or false alarm events triggered by the interference and noise. The MF based detector was proposed as an effective means of combating the interference, which is regarded as the optimal detector in additive white Gaussian noise (AWGN) environments [3]. However, the MF requires some prior knowledge of the primary signal to be detected, such as the pulse shape, modulation type and so on. As a further alternative, the feature extraction aided detector (e.g., cyclostationary FE) emerges as a promising sensing approach, which is capable of effectively distinguishing the primary signals from both the background noise and from the interference. This makes the FE detector robust to the background noise even in extremely low signal-to-noise ratio (SNR) scenarios. This advantage of the FE detector however comes at the cost of a high computational complexity, since it requires an extra training process for extracting the relevant signal features. Due to its high computational complexity, the FE detector consumes more energy than the ED and the MF detector. Table I provides a summary of the three types of detectors in terms of their robustness to noise variance uncertainty, the prior knowledge requirement, the computational complexity imposed, and the energy consumption.

Both the ED as well as the MF detector and the FE detector typically work well in Gaussian noise environments, but their receiver operating characteristics (ROC) severely degrade in wireless fading scenarios. {{Specifically, if a deep fade is encountered, the desired signal received at an unlicensed user may become too weak to be detected by the aforementioned ED, MF and FE detectors, thus causing a performance degradation.}} In order to combat the fading effects, cooperative spectrum sensing may be invoked [2] by allowing multiple users to cooperate with each other in identifying spectrum holes, where the multiple users first scan the licensed spectrum bands and then report their independent observations to a fusion center for the sake of making a final decision concerning the idle or busy status of the scanned spectral bands. It has been shown in [2] that cooperative spectrum sensing significantly outperforms the conventional non-cooperative approaches. This, however, comes at the expense of additional energy consumption, since the cooperative spectrum sensing consumes additional power during its reporting phase. To this end, another design alternative for spectrum sensing is to develop a geo-location incumbent database that keeps and periodically updates the information of licensed spectral occupancy in any given geographical location. In this way, unlicensed users can readily access the available unused spectrum by checking the incumbent database with their current geo-locations. This is an energy-efficient solution, but results in a degraded ROC performance, since the spectrum hole is determined using a propagation model of the primary signal, which fails to reflect real-world environments such as mountains, buildings, tunnels, and so on [4]. As a remedy, we can combine the cooperative sensing and geo-location database, aiming for striving a good compromise between the achievable performance and the energy savings.

\section{Exploiting Spectrum Holes with the Aid of Network Cooperation}
In this section, we consider the benefits of cooperation in efficiently exploiting the spectral access opportunities detected for the sake of energy conservation in green cellular networks. Fig. 1 shows a cellular network consisting of a BS and multiple UTs that support diverse wireless access interfaces, including both Bluetooth and Wi-Fi. Since the UTs (e.g., smart phones) are equipped with both Bluetooth and Wi-Fi, they are capable of establishing a secondary network within the cellular network and exploit the secondary network for assisting cellular communications to achieve energy efficiency improvements, where the secondary network operates within the spectrum holes detected. It is pointed out that there are some standardization efforts in developing Wi-Fi over the unoccupied TV bands, which is known as the IEEE 802.11af. This means that next-generation Wi-Fi and Bluetooth interfaces would be expected to operate in a wider range of spectral bands (including the TV bands), rather than being limited to the industrial, scientific and medical (ISM) bands only. If no spectrum holes are identified, the secondary network is deactivated and the UTs directly communicate with the BS within the cellular spectrum. By contrast, when a spectrum hole is detected with the aid of spectrum sensing, different wireless networks may be invoked for cooperating with classic cellular communications. This technique may thus be referred to as \emph{cognitive network cooperation}, since the above-mentioned cognitive feature was exploited.

\begin{figure}
  \centering
  {\includegraphics[scale=0.55]{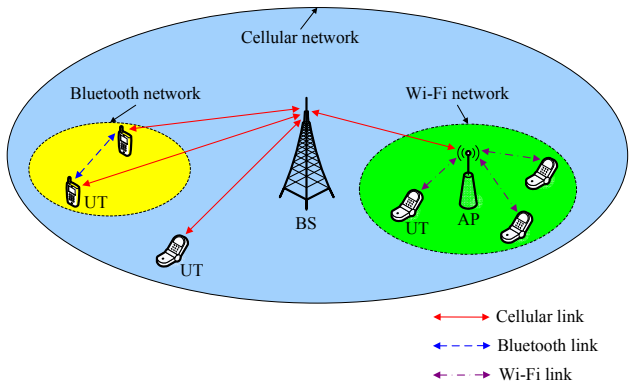}\\
  \caption{A heterogeneous wireless environment consisting of Bluetooth, Wi-Fi, and cellular networks.}\label{Fig1}}
\end{figure}
As shown in Fig. 1, when considering Wi-Fi as an example of the secondary network, the UTs first communicate with an access point (AP) through Wi-Fi links over the spectrum holes detected and then the AP exchanges data packets with the BS through cellular links over the cellular spectrum. In this way, the UTs can communicate with the BS relying on the AP for improving the attainable energy efficiency of cellular communications. As an alternative, we can also adopt Bluetooth for assisting cellular communications between the UTs and the BS, as depicted in Fig. 1. Without loss of generality, let us consider the cellular downlink transmissions from the BS to a pair of UTs, namely to U1 and U2. We first allow the BS to transmit its downlink data packets to U1 and U2, respectively, over the cellular spectrum. Due to the broadcast nature of the wireless medium, U1 (or U2) can overhear the transmissions from the BS to U2 (or U1). Then, if a spectrum hole is identified, U1 and U2 exchange their received signals through a two-way Bluetooth link over the spectrum hole detected. Therefore, the employment of Bluetooth over the spectrum holes provides spatial diversity for cellular communications and hence reduces the overall energy consumption under a specific target QoS requirement.

Naturally, a similar network cooperation process may be applied in cellular uplink transmissions from the UTs to the BS. To be specific, given that a spectrum hole is identified, we first allow the spatially-distributed UTs to communicate through a secondary network over the detected spectrum hole for exchanging their uplink packets. Once the UTs obtained each other's data packets, they can cooperate (with the aid of space-time coding) for the sake of transmitting their packets to the BS over the cellular spectrum. It can be observed that a secondary network is employed for assisting the cellular communications in an opportunistic manner, provided that a spectrum hole has been detected. This technique is more sophisticated than the traditional cooperative regime operating in a homogeneous network environment. Explicitly, in the traditional cooperation, a UT is required to transmit its data packet to its cooperative partner, which then forwards the received data to the BS, both transmissions occurring over the same cellular spectrum. This halves the cellular spectrum efficiency, since two orthogonal channels are needed for completing the transmissions from a UT to the BS via its partner. By contrast, intelligent cognitive network cooperation allows a UT to transmit its data to its partner using a secondary network over the detected spectrum hole, instead of using the cellular spectrum. This saves cellular spectrum resources by leveraging the idle spectrum hole and thus has the potential of improving the energy-efficiency as compared to the traditional cooperation.

\begin{figure}
  \centering
  {\includegraphics[scale=0.55]{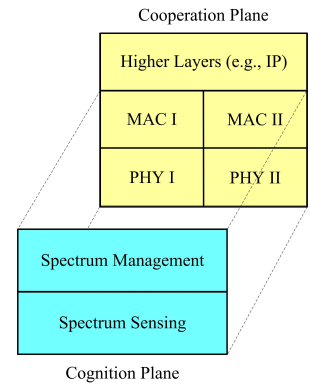}\\
  \caption{{Protocol reference model of the cognitive network cooperation.}}\label{Fig2}}
\end{figure}

In Fig. 2, we portray the protocol reference model of the proposed cognitive network cooperation composed of two planes, namely the cognition plane and the cooperation plane. As shown in Fig. 2, the cognition plane incorporates both spectrum sensing and spectrum management functions. The spectrum sensing functions are invoked for identifying spectrum holes, which are then allocated to the different network entities by the spectrum management function. Fig. 2 also illustrates an example of the cooperation plane, where two sets of medium access control (MAC) and physical layer (PHY) protocols are considered, which are denoted in the figure by PHY I-MAC I and PHY II-MAC II, respectively. It may be observed that the two PHY-MAC pairs (i.e., PHY I-MAC I and PHY II-MAC II) share common higher layers, namely common network (NET) and application (APP) layers. As a benefit, the different network access interfaces can be coordinated and controlled with the aid of the upper-layer protocols. It is worth mentioning that contemporary cellular devices (e.g., smartphones) typically support multiple PHY-MAC protocols for the sake of cooperating with different wireless access networks, including the 3G, LTE, Wi-Fi, Bluetooth and other networks. In practice, the energy consumed by the upper-layer protocol management is non-negligible and should be taken into account for minimizing the overall network energy dissipation. Thus, it is of particular interest to investigate the optimization of the cooperation plane by jointly considering the PHY, MAC and upper-layer protocols. {{Additionally, it is important to address the time scale and synchronization issues, when different network access interfaces cooperate with each other.}}

\section{Case Study: An Energy Efficient Cellular Communications Scheme Relying on Both Cognition and Cooperation}
In this section, a case study is provided for illustrating the attainable energy saving benefits of cognition and cooperation techniques in green cellular communications. Without loss of generality, let us assume that in a cellular network, the BS is transmitting its data packets to an intended UT over the cellular spectral band at a carrier frequency of $f_c$ and within the bandwidth of $B_c$. The cellular devices (both the BS and UTs) scan a TV channel using spectrum sensing and identify whether the scanned TV channel is idle or not, where the carrier frequency and the bandwidth of the TV channel are denoted by $f_t$ and $B_t$, respectively. If the TV channel is deemed to be idle, it can be utilized for cellular communications in many ways, including the enhancement of the transmission reliability and/or the improvement of the system throughput. For example, in the case of an idle TV channel, the BS repeats the transmission of its data packet to the UT over the TV channel by splitting its transmit power between the cellular and TV bands. Since these two frequencies are likely to be sufficiently far apart to experience independent fading, they are capable of providing a beneficial frequency diversity gain, hence improving the reliability and/or the energy efficiency, while maintaining the target QoS requirement.

\begin{figure}
  \centering
  {\includegraphics[scale=0.65]{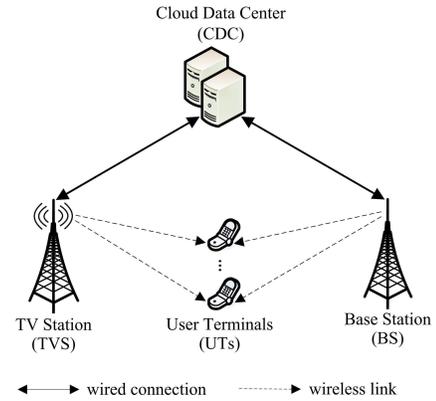}\\
  \caption{{{Joint cognition and cooperation aided cellular wireless networks.}}}\label{Fig3}}
\end{figure}
{{Fig. 3 shows the stylized illustration of joint cognition and cooperation aided cellular wireless networks, where the TV channel is utilized for assisting cellular transmissions from the BS to the UTs. Observe from Fig. 3 that the TVS and BS are assumed to be connected by a high-bandwidth optical backbone or microwave link, for example via cloud data center (CDC), which enables prompt data exchange between the TVS and BS. Once the TVS received the BS' data packets through the CDC, it can assist the BS' transmissions to the UTs by using Alamouti's space-time code [18], as shown in Fig. 4 (\emph{a}). More specifically, the BS' data packets are, respectively, denoted by $x_1$ and $x_2$, which are also known at the TVS via the CDC. According to Alamouti's space-time coding, the BS and TVS, respectively, transmit $x_1$ and $x_2$ to an intended UT at the same time over a cellular spectral band. Then, in the next time slot, they simultaneously transmit $-x^{*}_2$ and $x^{*}_1$ to the UT, where $*$ denotes the conjugate complex operation. This transmission requires cooperation between the BS and TVS, which operates over the cellular band, thus referred to as the cellular frequency plane as shown in Fig. 4 (\emph{a}). For fair comparison, the total power consumed at the BS and TVS is constrained to a fixed level denoted by $P$, namely to the power of a single transmission by the BS only. Moreover, an equal power-sharing is considered and thus the transmit powers of the BS and TVS are given by $P/2$. Additionally, if the TV channel is found to be idle, the BS and TVS may repeat the above-mentioned cooperative transmission over the idle TV channel, called TV frequency plane in Fig. 4 (\emph{a}). Finally, the UT applies Alamouti's decoding algorithm to recover the BS' data packets $x_1$ and $x_2$. Again, it can be observed in Fig. 4 (\emph{a}) that both the cognition and cooperation are invoked by the proposed transmission scheme.}}

\begin{figure*}
  \centering
  {\includegraphics[scale=0.8]{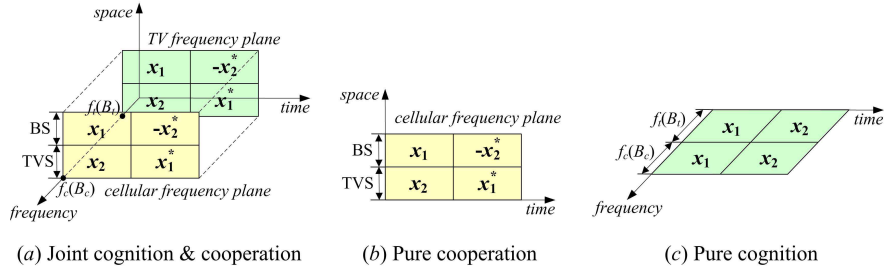}\\
  \caption{{{Time slot structures of different transmission schemes: (\emph{a}) the proposed joint cognition and cooperation, (\emph{b}) pure cooperation, and (\emph{c}) pure cognition.}}}\label{Fig4}}
\end{figure*}
For comparison, we also consider both the pure cooperation and pure cognition schemes, as shown in Figs. 4 (\emph{b}) and (\emph{c}), respectively. To be specific, the pure cooperation allows the TVS to assist the BS' transmission by using Alamouti's space-time code, where only the cellular frequency is utilized without exploiting the idle TV channel. By contrast, in the pure cognition scheme, only the BS transmits its data packets to its intended UT without relying on the TVS. As shown in Fig. 4 (\emph{c}), the BS transmits its packets $x_1$ and $x_2$ in consecutive time slots over the cellular band. If an idle TV channel is identified, the pure cognition scheme allows the BS to repeat its data transmission over the detected idle TV band. Naturally, due to the background noise and interference, it is impossible to achieve perfectly reliable spectrum sensing without missed detection and false alarm events. For notational convenience, the probabilities of detecting and false alarming the presence of TV subscribers, which are called successful detection probability and false alarm probability, are represented by $P_d$ and $P_f$, respectively. {{If a missed detection of the presence of TV subscribers occurs, the secondary cellular communications operating over the TV channel would interfere with the primary TV transmissions. In order to limit this mutual interference to a tolerable level, the detection probability $P_d$ and the false alarm probability $P_f$ should satisfy target values. In this paper, we consider $P_d = 0.99$ and $P_f = 0.01$, unless otherwise stated.

Additionally, let $P_a$ denote the probability that a TV channel is unoccupied by a TVS and hence becomes available for cellular communications. When the BS transmits its signal at the power of $P_{BS}$ and at a data rate of $R$, the power received by the UT, which is denoted by ${P_{R}}$, can be expressed as
\begin{equation}\label{equa1}
{P_{R}} = {P_{BS}}{\left(\dfrac{c }{{4\pi f_c d_{BU}}}\right)^2}{G_{BS}}{G_{UT}}|h_{BU}{|^2},
\end{equation}
where $c $ is the speed of light, $f_c$ is the cellular carrier frequency, $d_{BU}$ is the distance of the UT from the BS, ${G_{BS}}$ is the transmit antenna gain at the BS, ${G_{UT}}$ is the receive antenna gain at the UT, and $h_{BU}$ is the fading coefficient of the channel spanning from the BS to the UT. In this paper, the Rayleigh fading model is used for characterizing the channel, hence $|h_{BU}{|^2}$ obeys the exponential distribution with a mean of $\sigma _{BU}^2$. It is assumed that all receivers experience Gaussian distributed thermal noise with a zero mean and a variance of $\sigma _n^2$ that is modeled as $\sigma _n^2 = \kappa TB$, where $\kappa $ represents the Boltzmann constant (i.e., $\kappa  = 1.38 \times {10^{ - 23}}{\textrm{{ Joule/Kelvin}}}$), $T$ is the system temperature in $\textrm{{Kelvin}}$ and $B$ represents the channel bandwidth in $\textrm{{Hz}}$. Defining ${N_0} = \kappa T$ as the noise power spectral density (PSD) and considering a room temperature of $T = 290K$, we have ${N_0} =  - 174{\textrm{{ dBm/Hz}}}$. Using the data rate $R$ in $\textrm{{bits/s}}$ and the transmit power $P_{BS}$ in Watt, the energy efficiency expressed in Bits-per-Joule is defined as
\begin{equation}\label{equa2}
{\eta} = \dfrac{ R}{P_{BS}},
\end{equation}
which is used for evaluating the energy cost in terms of the number of bits delivered per Joule. It is observed from Eq. (2) that increasing the data rate $R$ (or decreasing the transmit power $P_{BS}$) improves the energy efficiency $\eta$ of cellular transmission. However, the outage performance of cellular transmission also degrades, when a higher data rate (or lower transmit power) is considered, since an outage event occurs more frequently upon increasing the data rate (or decreasing the transmit power). Therefore, there is a tradeoff between the energy efficiency and the outage probability. According to Shannon's coding theorem, an outage event occurs, when the channel capacity falls below the data rate. {{Observe that in the proposed joint cognition and cooperation scheme, the BS and TVS will repeat their cooperative transmission to the UT over the idle TV channel detected. Thus, using the law of total probability, the outage probability of the joint cognition and cooperation scheme can be formulated as}}
\begin{equation}\label{equa3}
\begin{split}
{P_{out}} = &\Pr \left( {{C^{\textrm{coop}}_{\textrm{cell}}} < R,{C^{\textrm{coop}}_{\textrm{TV}}} < R,\hat H = {H_0}} \right) \\
&+ \Pr \left( {{C^{\textrm{coop}}_{\textrm{cell}}} < R,\hat H = {H_1}} \right),
\end{split}
\end{equation}
{{where ${C^{\textrm{coop}}_{\textrm{cell}}}$ represents the channel capacity of the cooperative transmission operating over cellular spectrum band, ${C^{\textrm{coop}}_{\textrm{TV}}}$ is the channel capacity of the cooperative transmission over the TV spectrum band}}, $\hat H = {H_0}$ represents the event of an idle TV channel being detected, and $\hat H = {H_1}$ indicates that the TV channel is deemed to be occupied by a TVS. Again, owing to the background noise and interference, the spectrum sensing is prone to missed detection and to false alarm in the presence of a TVS. If a missed detection event happens, the TV users and UTs will interfere with each other, i.e. the interference may inflict an outage event on both the UT and on the TV receiver. For notational convenience, let $P_{TVS}$ and $d_{TU}$ represent the transmit power of the TVS and the distance of the UT from the TVS, respectively. Additionally, the wireless channel spanning from the TVS to UT is also modeled by the Rayleigh fading and the average fading-induced attenuation between the TVS and the UT is denoted by $\sigma^2_{TU}$. So far, we have conceptually characterized the relationship between the energy efficiency and outage probability of {{the proposed joint cognition and cooperation scheme. It is pointed out that we can similarly determine the energy efficiency and outage probability of the pure cooperation and pure cognition schemes of Figs. 4 (\emph{b}) and (\emph{c}).}}

\begin{table}
  \centering
  \caption{{System Parameters Used in Numerical Evaluation.}}
  {\includegraphics[scale=0.6]{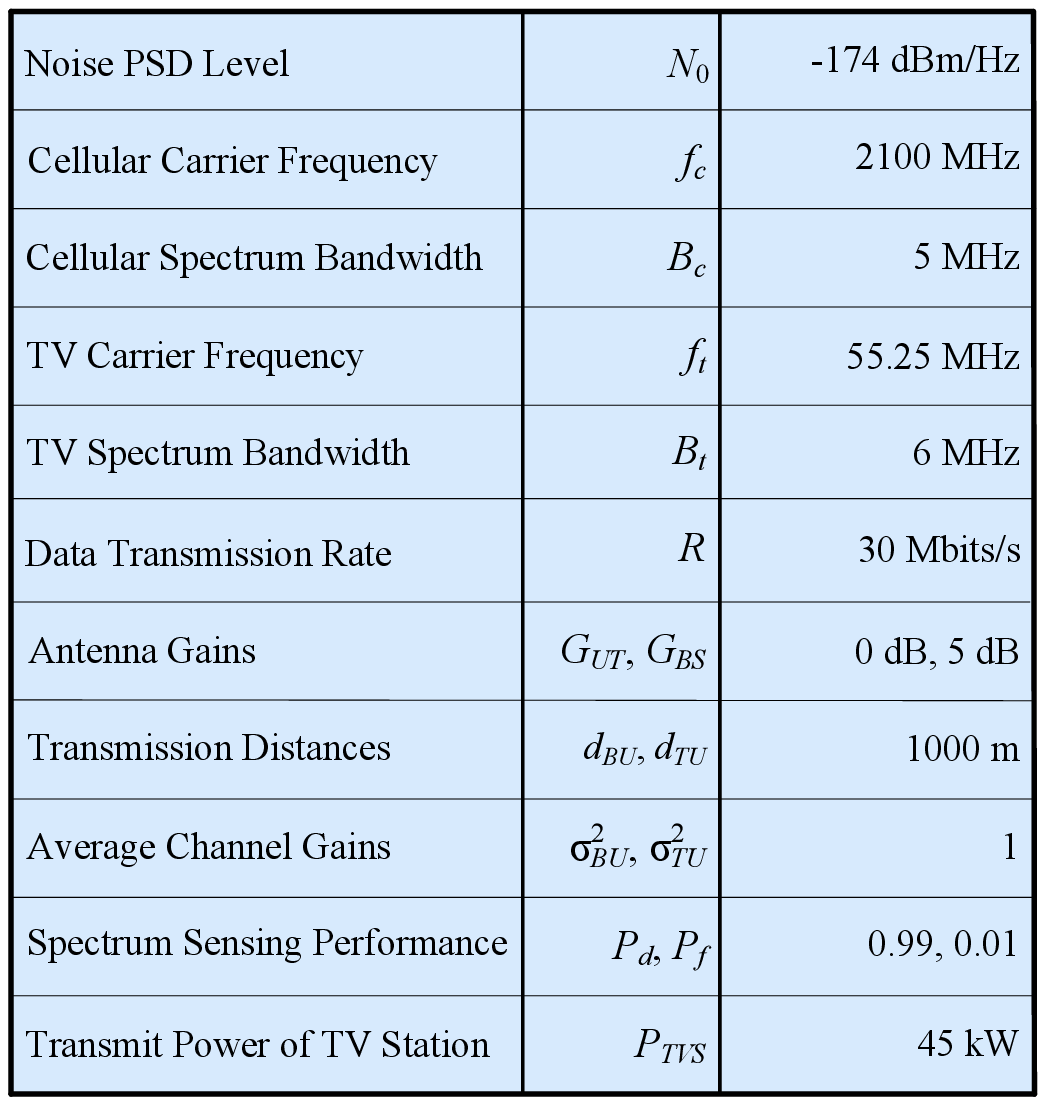}\label{Tab2}}
\end{table}
In what follows, we present some numerical results for characterizing the energy efficiency using Eqs. (1)-(3). Table II lists the system parameters used in our numerical evaluation, where ${f_c} = 2100{\textrm{ MHz}}$ and ${B_c} = 5{\textrm{ MHz}}$ correspond to the 3GPP LTE system. In North America, LTE operates at $2100 {\textrm{ {MHz}}}$ with various options of channel bandwidth, including $1.4 {\textrm{ {MHz}}}$, $3 {\textrm{ {MHz}}}$, $5 {\textrm{ {MHz}}}$, $10 {\textrm{ {MHz}}}$, $15 {\textrm{ {MHz}}}$ and $20 {\textrm{ {MHz}}}$. In the very high frequency (VHF) band ranging from $54 {\textrm{ {MHz}}}$ to $216 {\textrm{ {MHz}}}$, there are twelve TV channels, i.e., channels $2$-$13$. We consider the employment of TV channel $2$ having a carrier frequency of ${f_t}=55.25 {\textrm{ {MHz}}}$ and a bandwidth of ${B_t}=6 {\textrm{ {MHz}}}$ for our numerical evaluations. The data rate is given by $R=30{\textrm{ Mbits/s}}$ and the antenna gains of the UT and BS are specified as ${G_{UT}} = 0{\textrm{ dB}}$ and ${G_{BS}} = 5{\textrm{ dB}}$. The transmission distances spanning from the BS and TVS to the UT are $d_{BU}=1000{\textrm{ m}}$ and $d_{TU}=1000{\textrm{ m}}$, respectively, while the average fading gains are assumed to be $\sigma^2_{BU}=\sigma^2_{TU}=1$. Additionally, the successful detection probability and false alarm probability of $P_d=0.99$ and $P_f=0.01$ as well as a transmit power of $P_{TVS}=45\textrm{ kW}$ are considered in our numerical evaluation. Observe that in North America, the transmit power of digital TV stations in Band I (i.e., TV channels $2$-$6$) is typically limited to $45\textrm{ kW}$.

Fig. 5 shows the energy efficiency versus the outage probability of {{the traditional direct transmission, the pure cognition, the pure cooperation as well as the proposed joint cognition and cooperation schemes in conjunction with $P_a =0.8$, which is the probability that the TV channel is unoccupied by the TVS and hence becomes available for cellular communications. It is observed from Fig. 5 that upon increasing the outage probability, the energy efficiencies of the traditional direct transmission, the pure cognition, the pure cooperation as well as the proposed joint cognition and cooperation schemes are significantly improved. This explicitly shows that the energy efficiency of cellular communications improves as the outage performance degrades, implying the presence of a tradeoff between the energy efficiency and the outage probability.}}

\begin{figure}
  \centering
  {\includegraphics[scale=0.5]{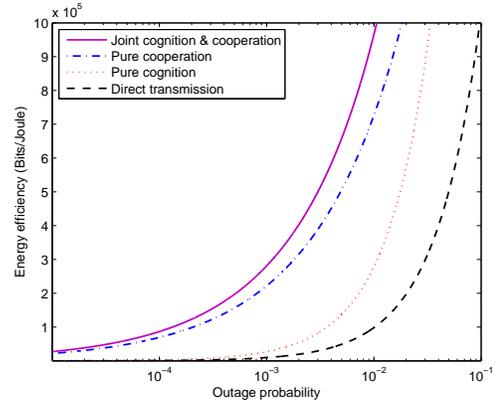}\\
  \caption{Energy efficiency versus outage probability of the traditional direct transmission, pure cognition, pure cooperation as well as the proposed joint cognition and cooperation schemes with $P_a =0.8$, where $P_a$ represents the probability that a TV channel is unoccupied by the TVS and becomes available for cellular communications.}\label{Fig5}}
\end{figure}
Fig. 5 also shows that across the entire outage probability region, the energy efficiency of the joint cognition and cooperation scheme is consistently higher than that of the traditional direct transmission, pure cognition and pure cooperation. In other words, given a maximum tolerable outage probability and a certain number of bits to be transmitted, the joint cognition and cooperation scheme dissipates less energy than the traditional direct transmission, pure cognition and pure cooperation approaches. This quantitatively demonstrates the energy saving benefits of exploiting the joint advantages of cognition and cooperation in cellular networks. Additionally, both the pure cooperation and the pure cognition achieve higher energy efficiencies than the direct transmission, showing that both the cooperation and cognition are beneficial in terms of energy consumption reduction. In view of the above case study, we may conclude that the energy dissipation of cellular networks can be significantly reduced by relying on both cognition and cooperation techniques.

\section{Conclusion and Future Work}
This paper studied the quantitative benefits of combined cognition and cooperation in the context of green cellular networks, where the cellular devices are first allowed to identify spectrum holes with the aid of their cognitive functionality (e.g., spectrum sensing) and then to cooperate for efficiently exploiting the spectrum holes detected to achieve energy savings. We summarized several spectrum hole identification approaches, including the low-complexity ED/MF based detection and FE detection. We then discussed their advantages and disadvantages in terms of their robustness to noise variance uncertainty as well as their computational complexity and energy efficiency. We proceeded by introducing the so-called cognitive network cooperation framework for the sake of the energy-efficient use of spectrum holes, where both the Bluetooth and Wi-Fi networks were invoked for supporting cellular communications. Finally, a quantitative case study of joint cognition and cooperation aided cellular communications was presented for characterizing the achievable energy savings in cellular communications. The results demonstrated that for a given outage probability and data rate, the total energy dissipated by cellular communications is significantly reduced by exploiting the joint cognition and cooperation.

{{In this paper, we only studied the energy efficiency of the proposed cognitive network cooperation without considering the energy overhead imposed by the coordination between different wireless network access interfaces. It is of high interest to build a testbed of the cognitive network cooperation and to critically appraise how the Bluetooth and Wi-Fi networks operate over the idle TV bands. Although our numerical results have demonstrated the energy efficiency advantage of the joint cognition and cooperation scheme over the conventional pure cognition and pure cooperation, this benefit has only been shown in theoretical studies relying on some idealized simplifying assumptions. It remains unknown, whether the joint cognition and cooperation approach is sufficient attractive in practical cellular networks in terms of reducing their energy consumption. We set aside this interesting problem for future research.}}

\begin{IEEEbiography}[{\includegraphics[width=1in,height=1.25in]{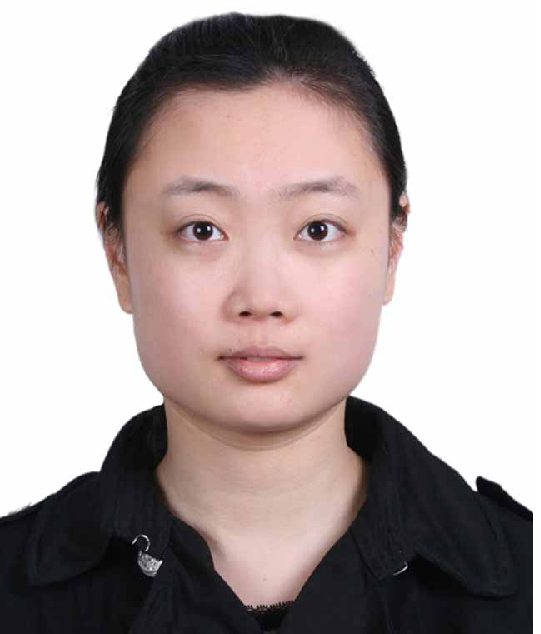}}]{Jia Zhu} is an Associate Professor at the Nanjing University of Posts and Telecommunications (NUPT), Nanjing, China. She received the B.Eng. degree in Computer Science and Technology from the Hohai University, Nanjing, China, in July 2005, and the Ph.D. degree in Signal and Information Processing from the Nanjing University of Posts and Telecommunications, Nanjing, China, in April 2010. From June 2010 to June 2012, she was a Postdoctoral Research Fellow at the Stevens Institute of Technology, New Jersey, the United States. Since November 2012, she has been a full-time faculty member with the Telecommunication and Information School of NUPT, Nanjing, China. Her general research interests include the cognitive radio, physical-layer security and communications theory.
\end{IEEEbiography}

\begin{IEEEbiography}[{\includegraphics[width=1in,height=1.25in]{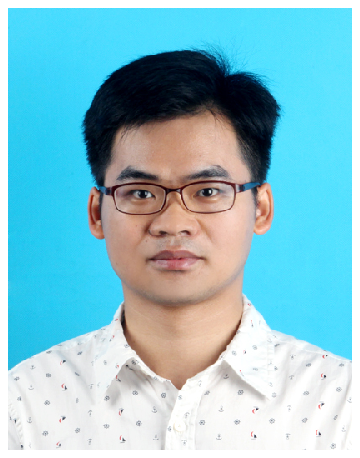}}]{Yulong Zou} (SM'13) is a Full Professor and Doctoral Supervisor at the Nanjing University of Posts and Telecommunications (NUPT), Nanjing, China. He received the B.Eng. degree in Information Engineering from NUPT, Nanjing, China, in July 2006, the first Ph.D. degree in Electrical Engineering from the Stevens Institute of Technology, New Jersey, the United States, in May 2012, and the second Ph.D. degree in Signal and Information Processing from NUPT, Nanjing, China, in July 2012. His research interests span a wide range of topics in wireless communications and signal processing, including the cooperative communications, cognitive radio, wireless security, and energy-efficient communications.

He was awarded the 9th IEEE Communications Society Asia-Pacific Best Young Researcher in 2014 and the co-recipient of Best (Student) Paper Award from the 80th IEEE Vehicular Technology Conference (IEEE VTC) in 2014. He has authored one Springer book entitled ``Physical-Layer Security for Cooperative Relay Networks". Dr. Zou is currently serving on editorial board for the IEEE Communications Surveys \& Tutorials, IEEE Communications Letters, IET Communications, EURASIP Journal on Advances in Signal Processing, China Communications, and KSII Transactions on Internet and Information Systems.

\end{IEEEbiography}

\end{document}